# Investigation of Numerical Diffusion in Aerodynamic Flow Simulations with Physics Informed Neural Networks

Alok Warey, Taeyoung Han, Shailendra Kaushik
Vehicle Systems Research Laboratory
General Motors Global Research and Development
Warren, MI 48090

---

## ABSTRACT

Computational Fluid Dynamics (CFD) simulations are used for many air flow simulations including road vehicle aerodynamics. Numerical diffusion occurs when local flow direction is not aligned with the mesh lines and when there is a non-zero gradient of the dependent variable in the direction normal to the streamline direction. It has been observed that typical numerical discretization schemes for the Navier-Stokes equations such as first order upwinding produce very accurate solutions without numerical diffusion when the mesh is aligned with the streamline directions. On the other hand, numerical diffusion is maximized when the streamline direction is at an angle of 45° relative to the mesh line. The amount of numerical diffusion can be reduced by mesh refinements such as aligning mesh lines along the local flow direction or by introducing higher order numerical schemes which may introduce potential numerical instability or additional computational cost. Couple test cases of a simple steady-state incompressible and inviscid air flow convection problem were used to investigate whether numerical diffusion occurs when using Physics Informed Neural Networks (PINNs) that rely on automatic differentiation as opposed to numerical techniques used in traditional Computational Fluid Dynamics (CFD) solvers. Numerical diffusion was not observed when PINNs were used to solve the Partial Differential Equation (PDE) for the simple convection problem irrespective of flow angle. The PINN correctly simulated the streamwise upwinding, which has great potential to improve the accuracy of Navier-Stokes solvers.

## 1. INTRODUCTION

Understanding flow phenomena and, especially, how aerodynamic forces are influenced by changes in vehicle body shape, are very important to improve vehicle aerodynamic performance particularly for low drag shapes. One of the goals of vehicle aerodynamics simulations is to predict the influence of changes in body shape on the flow field, and, thereby, on drag and lift forces. Due to numerical diffusion (numerical inaccuracy) of the current discretization schemes for the convection terms, it has been a challenge to predict accurate drag and lift forces with a reasonable number of mesh points. Therefore, the tendency is to increase the number of mesh points to improve numerical accuracy. However, the demand for mesh refinement is very high and typically a large number of mesh points (more than 100 million for external aerodynamics simulations) are needed. Typical external aerodynamic simulations with large number of mesh points become very inefficient due to slow turnaround times [1-5].

---

Corresponding author: Alok Warey (alok.warey@gm.com)

## 2. NUMERICAL DIFFUSION

Numerical diffusion (or inaccuracy) occurs when local flow direction is not aligned with the mesh lines and when there is a non-zero gradient of the dependent variable in the direction normal to the streamline direction. It has been observed that a typical numerical discretization scheme (first order upwinding) for the Navier-Stokes equations produces very accurate solutions without numerical diffusion when the mesh lines are aligned with the streamline directions. On the other hand, numerical diffusion is maximized when the streamline direction makes an angle of 45° with respect to the mesh line. The amount of numerical diffusion can be reduced by mesh refinements or by introducing higher order numerical schemes, which introduce potential numerical instability or additional computational cost [1-5].

To explain the effects of numerical diffusion, consider a simple steady-state convection problem for incompressible and inviscid air flow given by the energy equation below:

$$u\frac{\partial T}{\partial x} + v\frac{\partial T}{\partial y} = 0 \qquad [1]$$

The problem setup is shown in Fig. 1. Uniform velocity field was specified at the inlet boundary at different angles relative to the mesh line. A non-dimensional air temperature of 1.0 was specified above the dividing streamline and 0.0 below the dividing streamline. With no diffusion, no mixing layer should form and the exact solution for this equation should have a temperature discontinuity across the dividing streamline. If numerical diffusion is present in the calculations, then the numerical solution produces a mixing layer in the downstream flow. The amount of numerical diffusion introduced by the numerical scheme can be estimated by examining the size of the mixing zone across the dividing streamline.

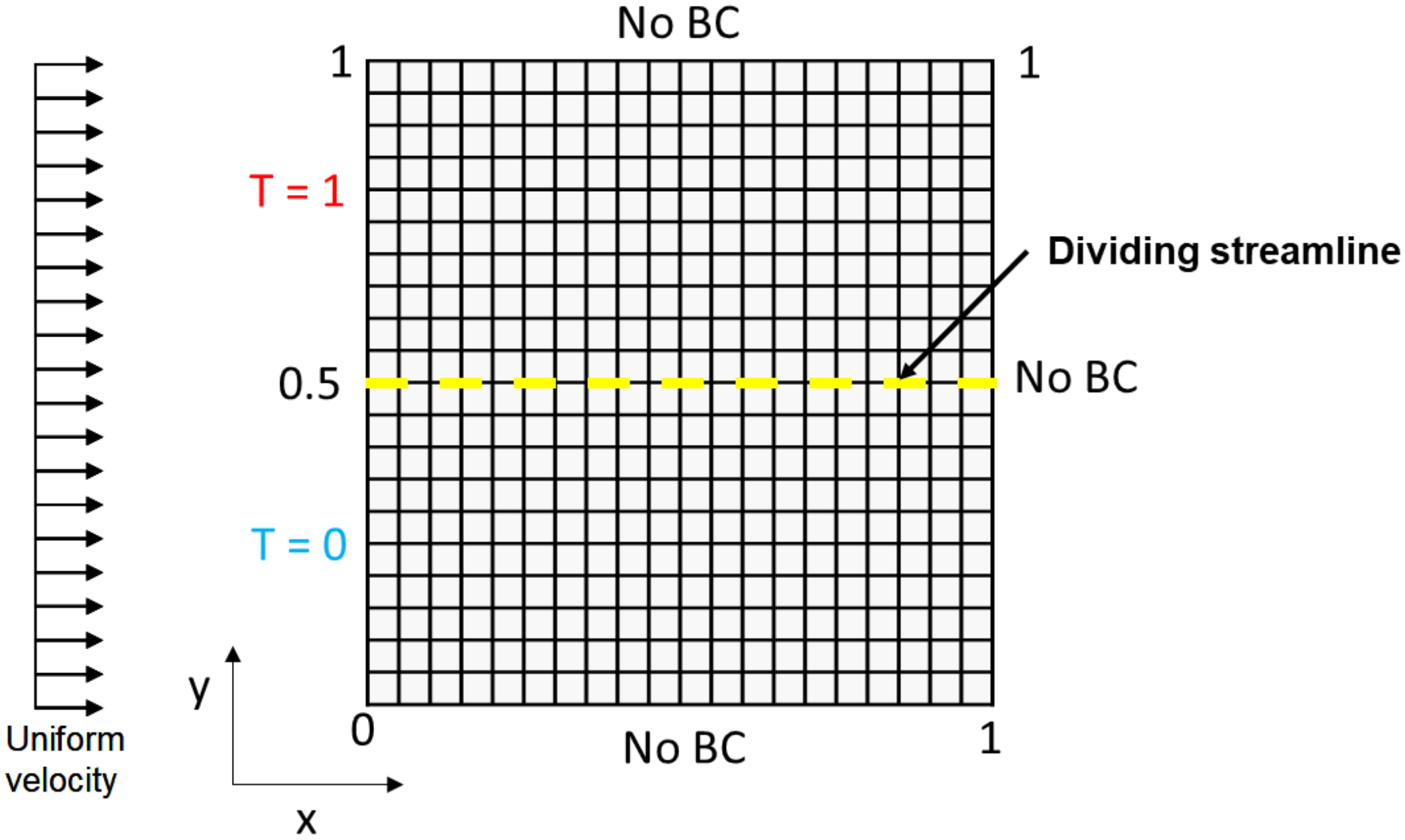

Fig. 1. Problem setup (Case 1).

Effect of numerical diffusion for one of the most popular numerical schemes in CFD, the first order upwinding scheme, was investigated for the following two cases:

- Case 1: Mesh lines are aligned with the flow direction (Fig. 1)
- Case 2: Flow angle relative to the mesh lines is 45° (Fig. 2)

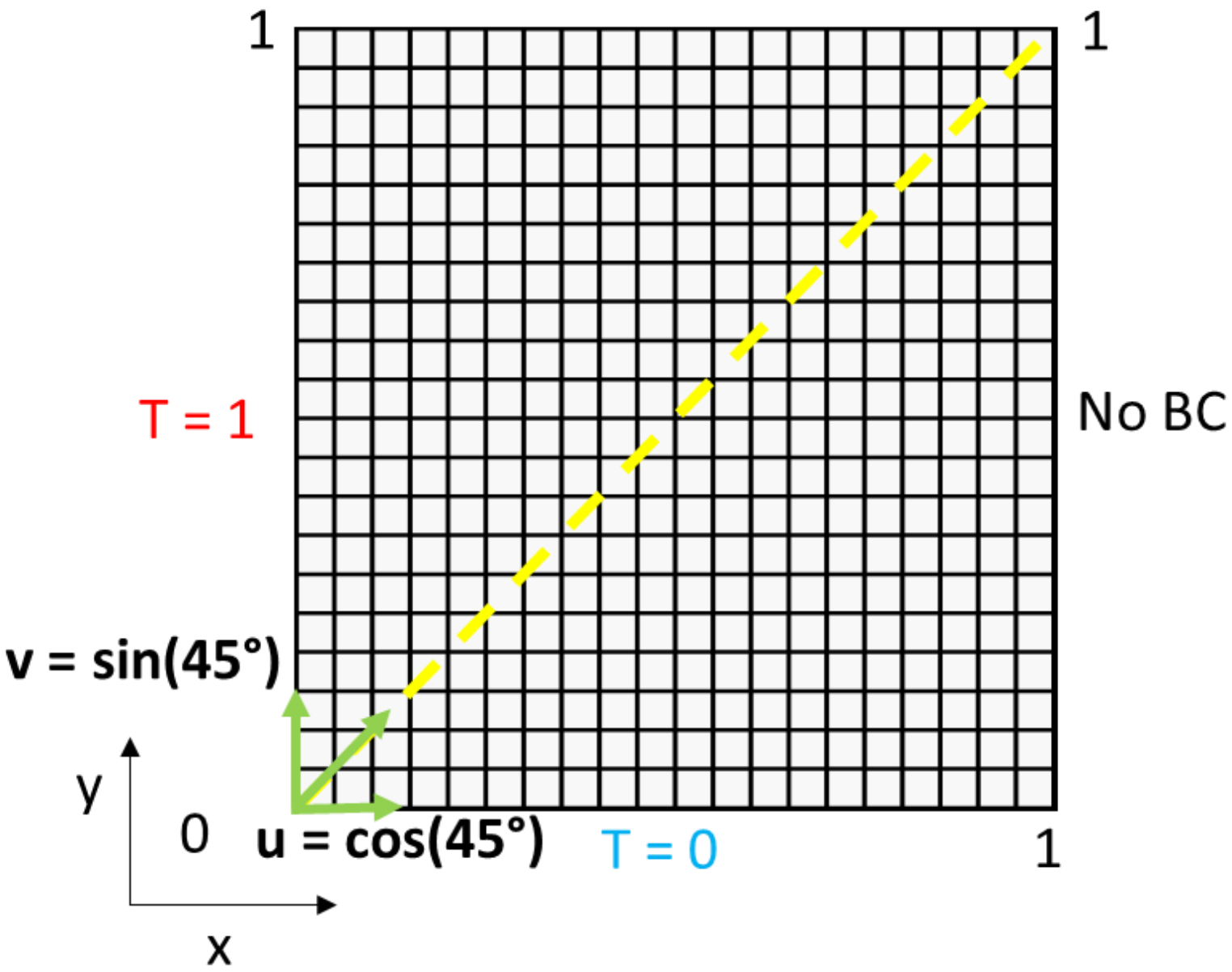

Fig. 2. Problem setup (Case 2).

Fig. 3 and 4 show the computed temperature field for both cases. As shown in Fig. 3, for Case 1, where the mesh lines are aligned with the flow direction, no mixing layer was observed, and a sharp temperature discontinuity persisted in the streamwise direction. For Case 2, significant numerical diffusion occurs along the 45° streamline as shown in Fig. 4.

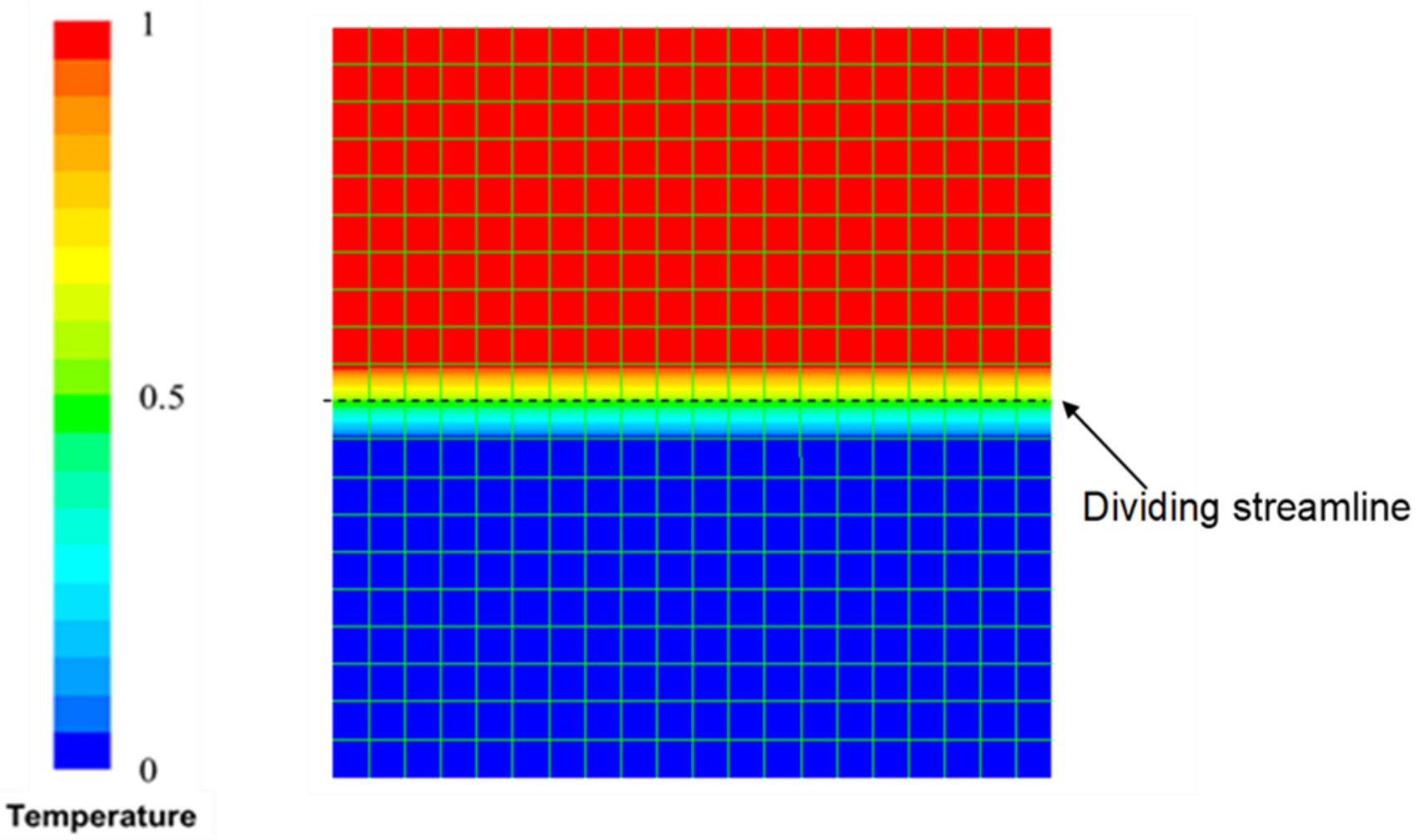

Fig. 3. Flow angle relative to the mesh line is 0°.

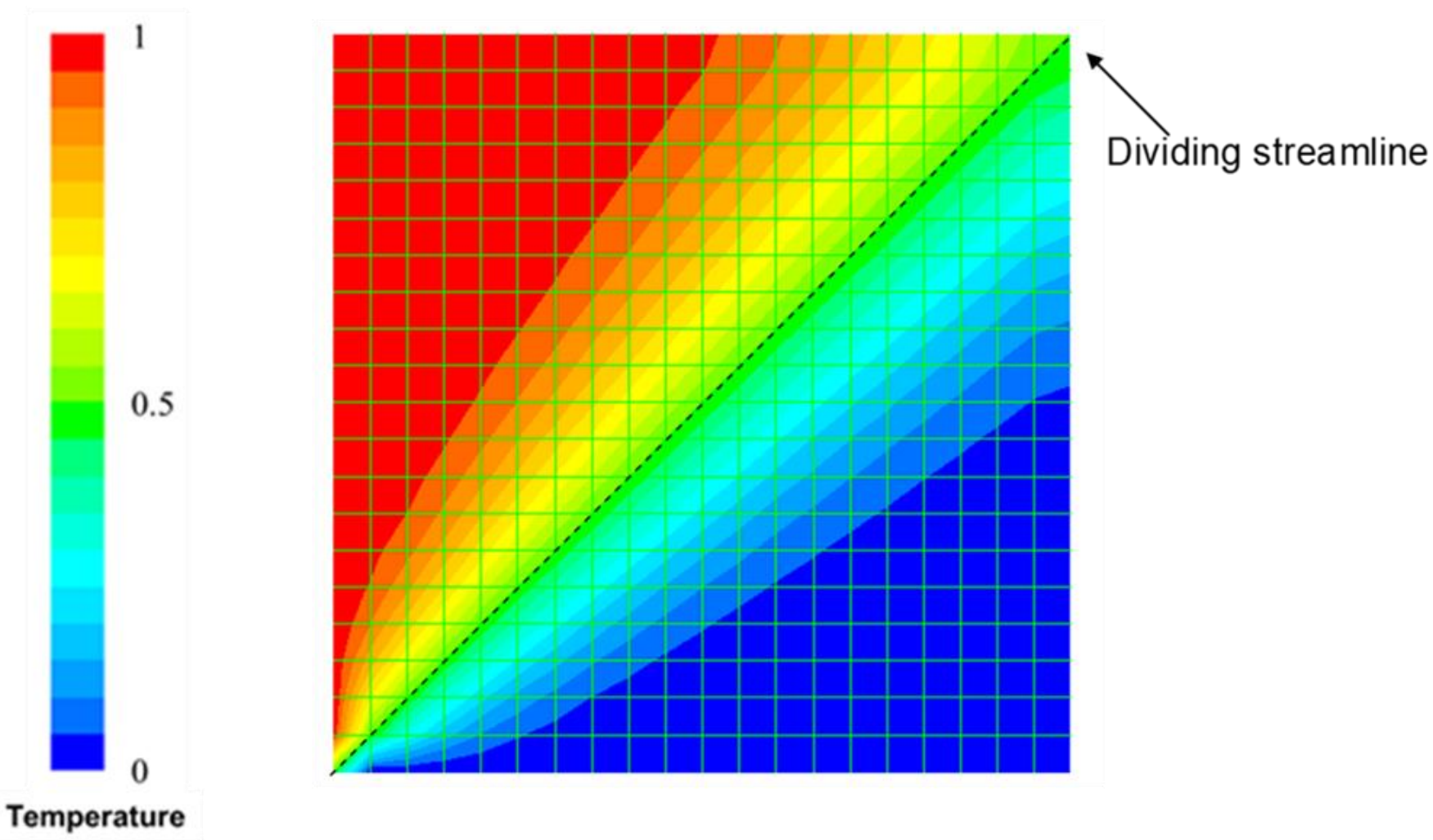

Fig. 4. Flow angle relative to the mesh line is 45°.

A significant amount of the numerical diffusion can be eliminated by simply orienting the mesh such that the mesh lines are more or less aligned with the streamline direction. As shown in Fig. 4 above, significant numerical diffusion is introduced in the solution when the flow direction is at 45° with respect to the mesh lines for the quad mesh. This is because no upstream flow information is readily available in the diagonal direction for the quad mesh. A simple approach to provide the upstream information in the diagonal direction is to introduce mesh lines in the diagonal direction by splitting the quad elements in the diagonal direction. As shown in Fig. 5, no mixing layer was formed, and the temperature discontinuity persisted in the diagonal direction. This clearly demonstrates the accuracy of the numerical approximation even for the first order upwinding scheme if one of the mesh lines is aligned in the direction of the streamline [5].

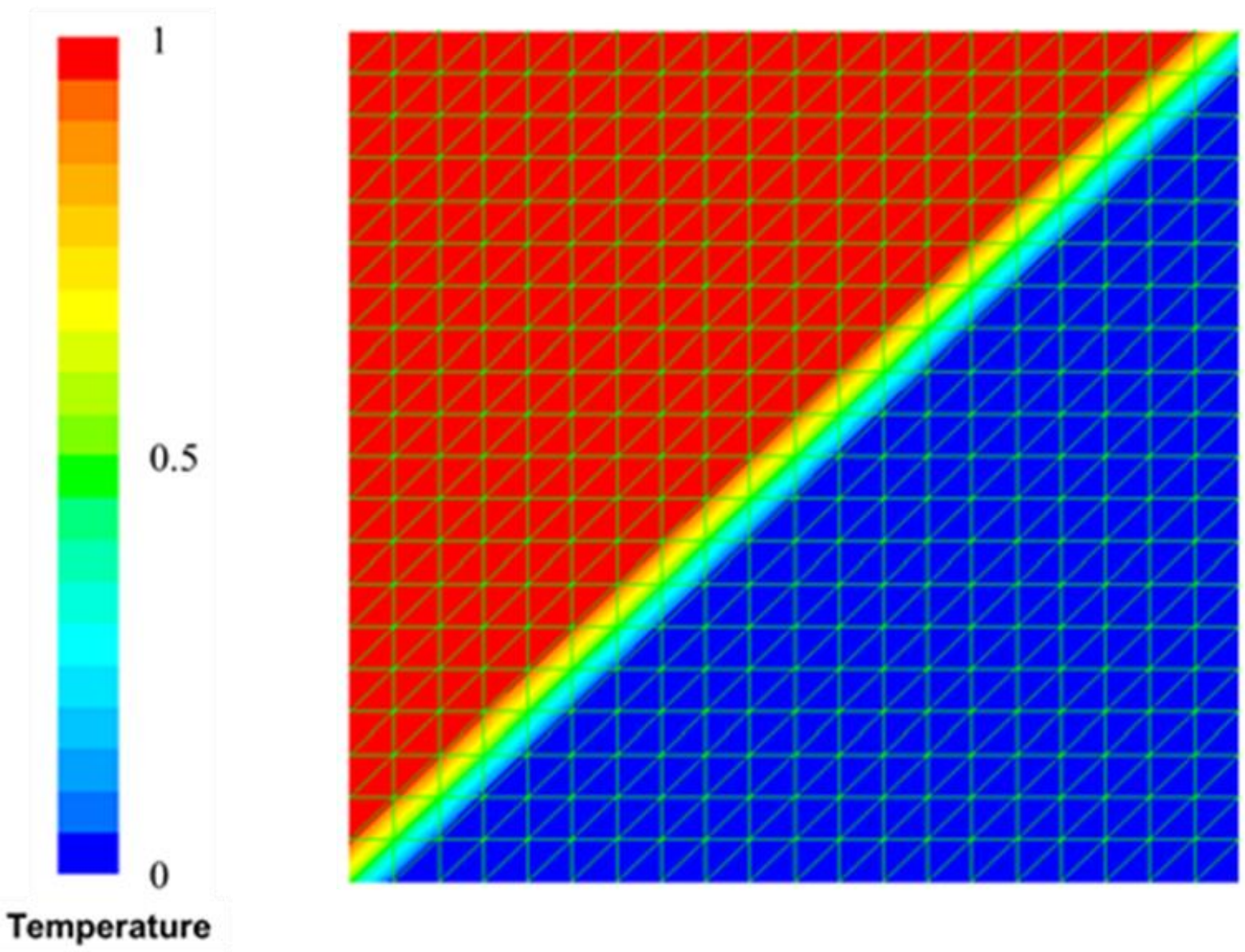

Fig. 5. Simple convection problem with the mesh adaptation to the flow direction (45°).

A three-dimensional boundary layer on the vehicle surface has a cross-flow velocity component in viscous flow regions. Therefore, aligning the mesh lines along the local flow direction within the boundary layers is not a trivial exercise. However, the effect of numerical diffusion in the boundary layer becomes less severe as viscous diffusion starts dominating in the boundary layer [5].

## 3. PHYSICS INFORMED NEURAL NETWORKS

Scientific Machine Learning (SciML) is an emerging research area focused on machine learning in the context of complex applications across science and engineering. Neural networks can be used as a method for efficiently solving difficult partial differential equations (PDEs) that are commonly encountered in science and engineering - Physics-Informed Neural Networks (PINNs). Physics is explicitly imposed by constraining the output of conventional neural network architectures. PINNs provide a mesh free alternative compared to traditional numerical methods and the potential to significantly reduce computational costs [6-8]. A schematic of a physics informed neural network for solving equation 1 above is shown in Fig. 6.

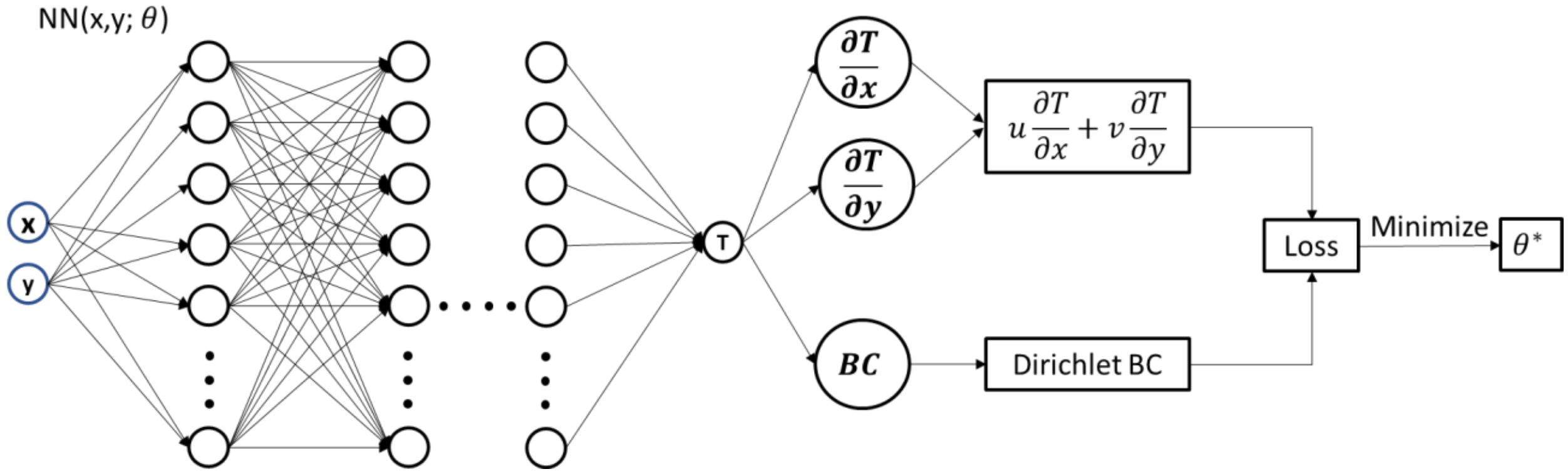

Fig. 6. Schematic of the PINN used to solve equation 1.

Cases 1 and 2 described above were used to investigate whether numerical diffusion occurs when using PINNs that rely on automatic differentiation as opposed to finite differencing. The PINN was implemented with the Python library, DeepXDE [8], which is designed to serve both as an education tool as well as a research tool for solving problems in computational science and engineering. A uniform grid of 5041 points in the domain (71x71 grid) and 1000 points along the boundary were used for training the PINN. Predictions with the trained network were done on a 200x200 grid.

Fig. 7 shows the predicted temperature field on a 200x200 grid by a trained PINN for Case 1: flow direction = 0°. No mixing layer was observed, and a sharp temperature discontinuity persisted in the streamwise direction similar to the CFD solution.

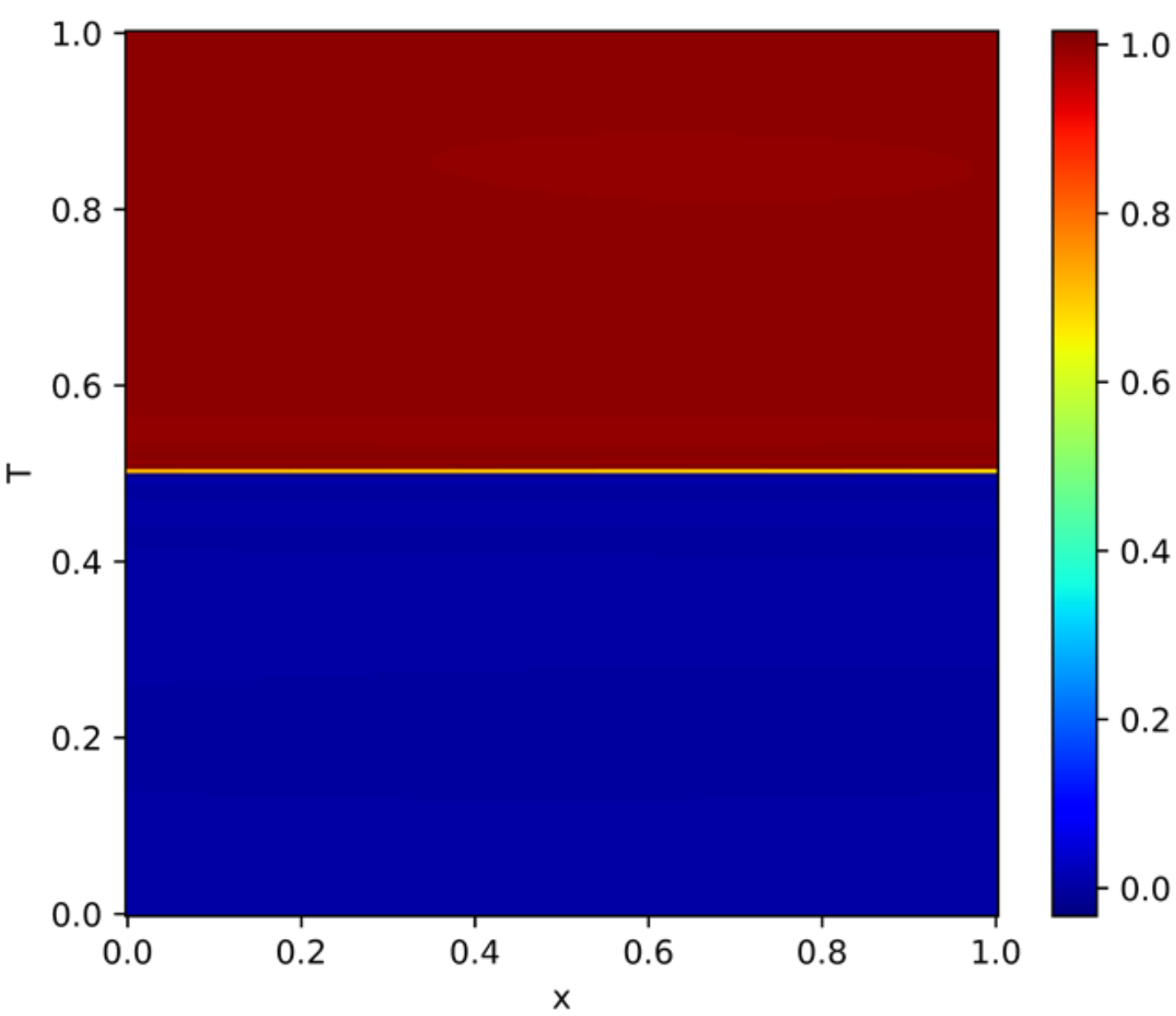

Fig. 7. Predicted temperature field on a 200x200 grid by a trained PINN for Case 1: flow direction = 0°.

Fig. 8 shows the predicted temperature field on a 200x200 grid by a trained PINN for Case 2: flow angle = 45°. No numerical diffusion was observed unlike the first order upwinding numerical scheme (Fig. 4).

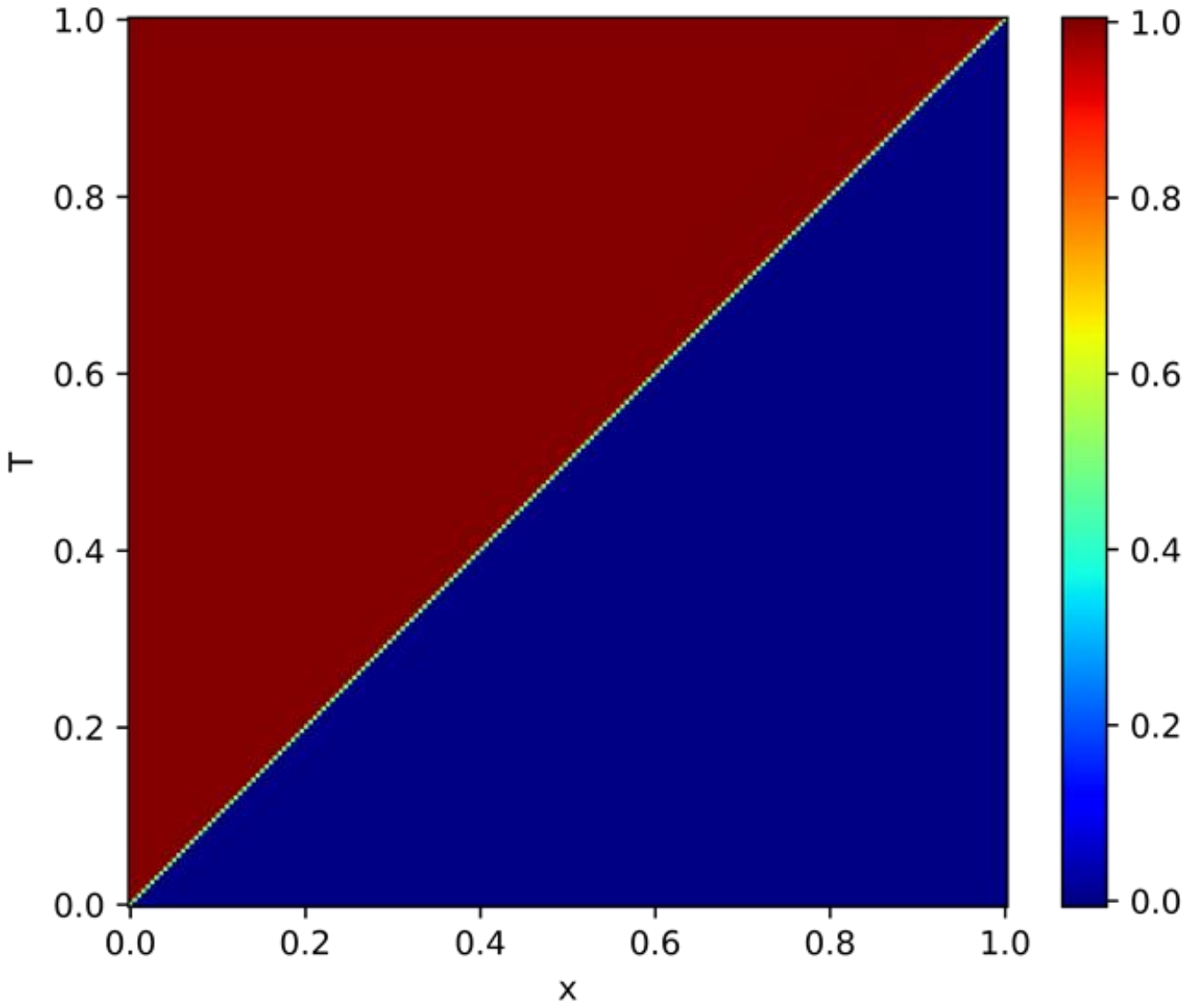

Fig. 8. Predicted temperature field on a 200x200 grid by a trained PINN for Case 2: flow angle = 45°.

Traditional discretization schemes such as finite-difference, finite-volume or even finite-elements used in CFD simulations use Taylor-series expansion for gradients. Number of terms retained in the discretization scheme (i.e. the size of the discretization stencil) determines the accuracy of the scheme or the truncation error. For example, the first order upwinding scheme for spatial derivatives introduces an error term $\frac{\partial^2 u}{\partial x^2}$ – which is essentially the viscosity term. However, the Taylor series analysis reveals that the truncation/discretization errors of the differential equations is not directly related with the numerical diffusion as shown in Fig. 3. Significant numerical diffusion appears in multi-dimensional flows when the differencing scheme fails to account for the true direction of the flow. In the present study, numerical errors associated with numerical diffusion were investigated for a simple test problem as shown in Fig. 9 and 10. Severe errors are introduced when the flow angle is not aligned with mesh lines as shown in Fig. 9. Practically, it is impossible to align the mesh lines to the streamlines in three dimensional (3D) flows. To reduce numerical diffusion, higher-order numerical schemes are applied. However, these schemes are not bounded and can be potentially unstable. Higher-order numerical schemes often introduce artificial diffusion to stabilize the solution. The PINN approach does not need an artificial viscosity to stabilize the solution with no truncation errors in the gradient calculations. The real advantage of PINN seems to be that it naturally adapts the streamwise upwinding, which is the basic characteristic of pure convection. For the test problem with 30° flow angle, there is no upstream temperature information for a uniform point distribution in the computational domain, although, for 0° and 45° flow angles, upstream flow information is available from points upstream. The second order upwinding scheme tends to reduce the numerical diffusion as shown in Fig. 10, however, the PINN demonstrated far better accuracy for various flow angles. The PINN correctly simulated the streamwise upwinding, which has great potential to improve the accuracy of Navier-Stokes solvers. A numerical scheme must satisfy the necessary criteria for a successful solution of convection-diffusion formulations. All of this can be completely circumvented by using PINN's and automatic differentiation.

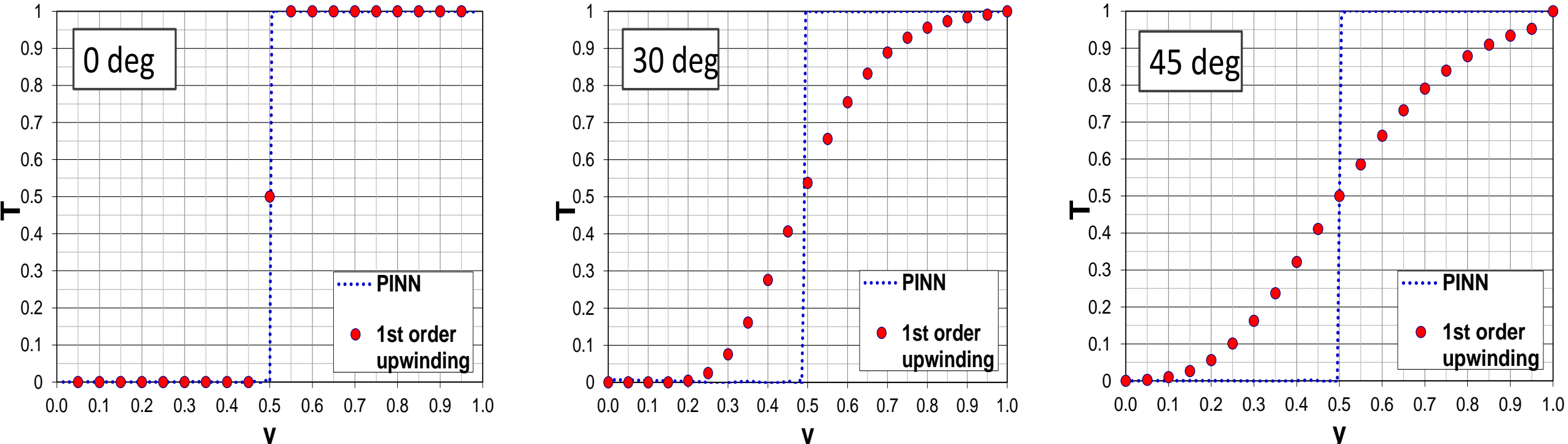

Fig. 9. Comparison of the temperature profile at x=0.5 for various flow angles = 0°, 30°, 45° between 1st order upwinding and PINN.

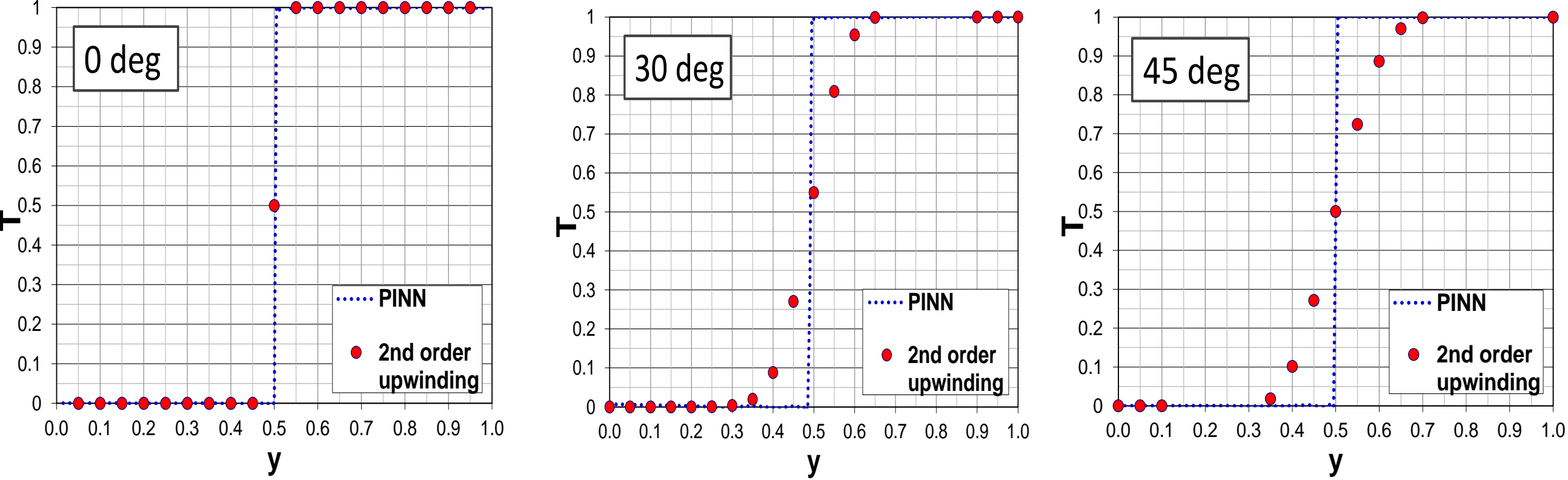

Fig. 10. Comparison of the temperature profile at x=0.5 for various flow angles = 0°, 30°, 45° between 2nd order upwinding and PINN.

## SUMMARY

Numerical diffusion (or inaccuracy) occurs with typical numerical schemes for the Navier-Stokes equations such as first order upwinding. The amount of the numerical diffusion can be reduced by mesh refinements such as aligning mesh lines along the local flow direction or by introducing higher order numerical schemes. Couple of test cases of a simple steady-state convection problem for incompressible and inviscid air flow were used to investigate whether numerical diffusion occurs when using Physics Informed Neural Networks (PINNs) that rely on automatic differentiation as opposed to numerical techniques used in traditional Computational Fluid Dynamics (CFD) simulations. The PINN correctly simulated the streamwise upwinding, which has great potential to improve the accuracy of Navier-Stokes solvers. A numerical scheme must satisfy the necessary criteria for a successful solution of convection-diffusion formulations. All of this can be completely circumvented by using PINN's and automatic differentiation.